\documentclass[12pt]{iopart}
\usepackage{graphicx}
\usepackage{dcolumn}
\usepackage{bm}
\usepackage{bbm}
\usepackage{multirow}
\usepackage{bbm}
\usepackage{graphicx}
\usepackage{float}
\usepackage{cases}

\begin{document}

\title[]{Security flaw of counterfactual quantum cryptography in practical setting}

\author{Yan-Bing Li$^{1,2,3}$, \quad Qiao-yan Wen$^{1}$, \quad Zi-Chen Li$^{2}$}

\address{
$^{1}$State Key Laboratory of Networking and Switching Technology, Beijing University of Posts and Telecommunications, Beijing, 100876, China\\
$^{2}$Beijing Electronic Science and Technology Institute,Beijing
100070,China\\
$^{3}$Department of Electrical Engineering and Computer Science,
Northwestern University, Evanston, Illinois 60208, USA}
\ead{liyanbing1981@gmail.com}

\begin{abstract}
Recently, counterfactual quantum cryptography proposed by T. G.
Noh [Phys. Rev. Lett. 103, 230501 (2009)] becomes an interesting
direction in quantum cryptography, and has been realized by some
researchers (such as Y. Liu et al's [Phys. Rev. Lett. 109, 030501
(2012)]). However, we find out that it is insecure in practical
high lossy channel setting. We analyze the secret key rates in
lossy channel under a polarization-splitting-measurement attack.
Analysis indicates that the protocol is insecure when the loss
rate of the one-way channel exceeds $50\%$.
\end{abstract}

\maketitle

\section{\label{sec:level1}Introduction}

Quantum cryptography allows higher security than classical
cryptography as it is based on the laws of physics instead of the
difficulty of solving mathematical problems. Quantum key
distribution (QKD)\cite{bb84}-\cite{y00}, which is to provide
secure means of distributing secret keys between the sender
(Alice) and the receiver (Bob), is often used to represent quantum
cryptography as the primary most important part. Now it has been
researched and developed in both theoretics and experiments. In
theoretic, QKD could offer unconditional security guaranteed by
the laws of physics\cite{lc99}. But due to the limitations of
real-life setting\cite{bl00}, such as the imperfect source,
imperfect detector, loss and noise in channel, practical QKD has
security loopholes and has suffered some attacks, such as photon
number splitting (PNS) attack\cite{hi95}, Trojan-horse
attack\cite{gf06}, faked state attack\cite{ma06}. On the other
hand, some achievements, such as decoy states mothod\cite{h03},
measurement-device-independent QKD (MDI-QKD) scheme\cite{lc12}
were made to let practical QKD be more secure.

Recently, counterfactual quantum cryptography proposed by
Noh\cite{n09} has attracted a lot of research, which allows
participants to share secret information using counterfactual
quantum phenomena. It is believed that the security is based on
that quantum particles carrying secret information are seemingly
not transmitted through quantum channels. So far, some security
proof\cite{yc10}, improvements\cite{sw10} and experimental
demonstrations\cite{rw11}-\cite{lj12} of counterfactual quantum
cryptography have been proposed.

However, we find out the counterfactual quantum
cryptography\cite{n09} is insecure in practical long distance
communication. The secret key rate will be $0$ under a
polarization-splitting-measurement attack when the loss rate of
the one-way channel is no less than $50\%$. Namely, the
eavesdropper (commonly called Eve) can obtain all the secret
information. Nevertheless, the cheat is unknowable to Alice and
Bob because its effect just likes a reasonable loss in practical
channel.

This paper is organized as follows. Sec. II reviews the
counterfactual QKD proposed in Ref.\cite{n09}. In Sec. III, We
analyze the error rate of raw key in lossy channel. A
polarization-splitting-measurement attack is given in Sec. IV. In
Sec. V, we analyze the secret key rate under the attack. Finally,
a short conclusion are provided in Section VI.

\section{\label{sec:level1}Counterfactual QKD}

\begin{figure}[h]
\centering
\includegraphics[scale=0.8]{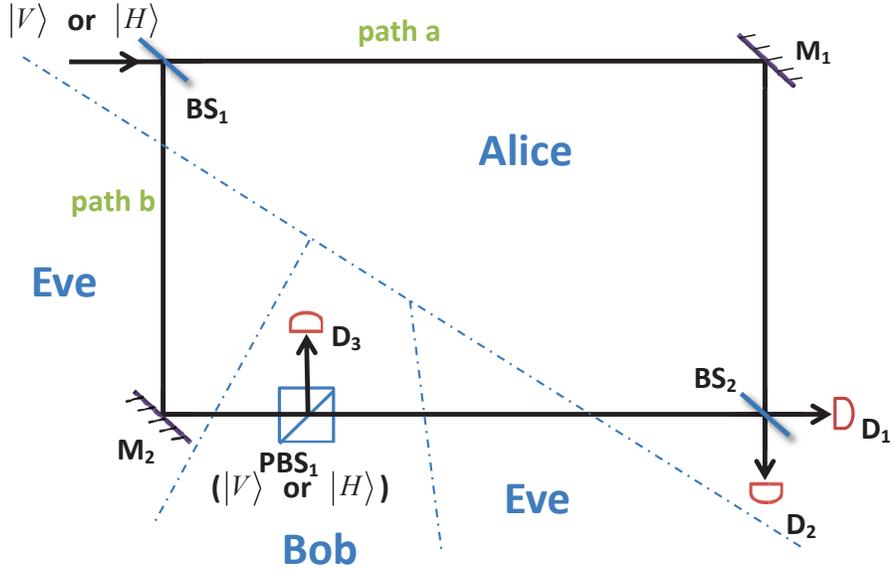}
\caption{(color online).\space The schematic of counterfactual
QKD. For simpleness, we have made some equivalent adjustments on
the original one. Whole space is divided by dotted line into three
sub-spaces, Alice's site, Bob's site and  public space (i.e.,
Eve's space). Alice sends the $i$th single-photon in state
$|V\rangle$ or $|H\rangle$, representing bit $0$ or $1$, to beam
splitter $BS_1$. Then the split pulses are transmitted into two
paths $a$ which is always in Alice's site, and $b$ which is in
public space toward Bob's site. Bob randomly uses $|V\rangle$
(representing $0$) or $|H\rangle$ (representing $1$) $PBS$ to
block the pulse in path $b$ when his bit is identical to Alice's,
or let it pass when his bit is differ to Alice's. When their bits
are different, detectors $D_2$ should always click since the
interferometry happens in $BS_2$. Else when their bits are same,
the detectors $D_1$ and $D_2$ and $D_3$ will click with some
probabilities since interaction-free measurement happens.
Additional, it is assumed that all of $D_1$, $D_2$ and $D_3$ could
detect the state's polarization $|V\rangle$ or $|H\rangle$. All
the $D_2$'s and $D_3$'s clicks and a part of $D_1$'s clicks are
used to detect eavesdropping, and the rest of $D_1$'s clicks with
correct polarization are used as the raw key.}
\end{figure}
Fig.1 is the schematic of counterfactual QKD\cite{n09}. For
simpleness, we have made some equivalent adjustments on it. Alice
triggers the single-photon source $S$, which emits a short optical
pulse containing a single photon at a certain time interval. She
randomly chooses the photon polarization in $|V\rangle$
representing the bit value ¡°0¡±, or $|H\rangle$ representing the
bit value ¡°1¡±. Thereafter, the photon enters a beam splitter
$BS_1$ and is split to two wave pulses $s_a$ and $s_b$. Then the
system state evolves into one of the following states:
\begin{subequations}
\begin{eqnarray}
|\phi_0\rangle=
 \sqrt{R}|0\rangle_a|V\rangle_b+\sqrt{T}|V\rangle_a|0\rangle_b,
\end{eqnarray}
\begin{eqnarray}
 |\phi_1\rangle=
 \sqrt{R}|0\rangle_a|H\rangle_b+\sqrt{T}|H\rangle_a|0\rangle_b.
\end{eqnarray}
\end{subequations}
where subscripts $a$ and $b$ represent the path towards Alice's
site and the path toward Bob's site, respectively, and $|0\rangle$
denotes the vacuum state in the path $a$ or $b$. $R$ and $T=1-R$
are the reflectivity and transmissivity of both $BS_1$ and $BS_2$,
respectively.

Bob has two polarizing beam splitter (PBS), $|V\rangle$ PBS
(representing the bit value ¡°0¡±) and $|H\rangle$ PBS
(representing the bit value ¡°1¡±), where $|V\rangle$ (or
$|H\rangle$) PBS means it addresses the state $|V\rangle$ (or
$|H\rangle$) towards detector $D_3$, while the state $|H\rangle$
(or $|V\rangle$) is sent towards the beam splitter $BS_2$. He
randomly chooses to use the $|V\rangle$ PBS or $|H\rangle$ PBS as
his device $PBS_1$.

If Alice and Bob's bits are different, the pulse on path $b$ will
be reflected by Bob and combined again at Alice's device $BS_2$.
The case just likes an interferometry with a single photon. In the
ideal setting, detector $D_2$ will click with certainty. Else if
Alice and Bob's bits are identical, the path $b$ will be blocked
by Bob's $PBS_1$. The case just likes an interaction-free
measurement with a single photon. Here the state $|\phi_0\rangle$
will be collapsed to $|0\rangle_a|V\rangle_b$ or
$|V\rangle_a|0\rangle_b$, $|\phi_1\rangle$ will be collapsed to
$|0\rangle_a|H\rangle_b$ or $|H\rangle_a|0\rangle_b$. In the ideal
setting, detector $D_1$, $D_2$ and $D_3$ will click with
probability $RT$, $T^2$ and $R$, respectively.

So in the ideal setting, $D_1$ clicks means Alice's source photon
basis and Bob's PBS basis are identify. Then Alice and Bob have a
certain amount of identify bits, some of which could be used to
check possible eavesdropping, and the rest of with could be used
as raw key bits. And some statistical laws are between $D_2$'s,
$D_3$'s clicks and Alice, Bob's bits, which could be used to check
possible eavesdropping and judge error rate. Additional, it is
assumed that all the detectors $D_1$, $D_2$ and $D_3$ could detect
the state's polarization $|V\rangle$ and $|H\rangle$, which also
could be used to check possible eavesdropping.

Since the raw key bits come from the events of $D_1$ clicks which
means that Bob's measurement result is vacuum state, peoples feel
that the participles which carry secret information seemingly have
not travelled between Alice and Bob. In fact, its security is
based on a type of noncloning principle for orthogonal
states\cite{n09}: if reduced density matrices of an available
subsystem are nonorthogonal and the other subsystem is not allowed
access, it is impossible to distinguish two orthogonal quantum
states $|\phi_0\rangle$ and $|\phi_1\rangle$ without disturbing
them.

\section{\label{sec:level1}Users' error raw key rate depend on lossy channel}

Similar to other QKD schemes, the limitations of real-life setting
will also bring some troubles to counterfactual QKD. Specially,
high lossy channel will be a formidable difficulty to it. In this
section, we will analyze the error rate of users' raw key pair,
i.e., the different rate of Alice and Bob's raw key pair, depend
on lossy channel. ( Besides the the loss in channel, some other
loss also appear in the source and the devices and some noise
appear in the source, channel and the devices, but they are not in
the paper's range.)

In the counterfactual QKD, the raw key rate is  proportional to
the single detector click rate ( i.e., the rate of the case in
which only one of detectors $D_1$, $D_2$ and $D_3$ clicks), which
will be affected by source single photon rate $R_{single}$ and the
loss rate. Symmetrically, we suppose that both the loss rates in
channel from Bob to Alice, and that from Alice to Bob are $\eta$,
i.e., the single photon will loss with probability $\eta$ in one
of the two channels. We recall that (1) Alice's raw key bits are
generated from the source single photons' bases. (2) Bob's raw key
bits are generated from his $PBS_1$'s basis, i.e., state in which
basis would be sent from $PBS_1$ toward $D_3$.

Then we analyze the cases in which the raw key will be generated
by Alice and Bob. The analysis will be done on one single photon
sent by Alice, which is in state $|V\rangle$ or $|H\rangle$ with
probability $1/2$ respectively. And we suppose that the channel
loss in the channel in public space and Bob's site, which is
denoted as channel $c_{A\rightarrow B\rightarrow A}$ and could be
divided to two parts ( the channels from Alice to Bob
$b_{A\rightarrow B}$ and from Bob to Alice $b_{B\rightarrow A}$),
is independent with state's polarization $|V\rangle$ and
$|H\rangle$, i.e., all the possible wave pulse will loss when
channel loss happens. ( Note that the channel loss in
$c_{A\rightarrow B\rightarrow A}$ is different to the loss happens
in Bob's PBS in which only one polarization is blocked. And also
note that channel loss in $c_{A\rightarrow B\rightarrow A}$ does
not mean that the photon vanishes in $c_{A\rightarrow B\rightarrow
A}$ with certainly since it might go through path $a$ probably.)
Theses cases are divided by two elements (i) whether the loss
happens or not in the channel in public space and Bob's site (then
we divide the channel $c_{A\rightarrow B\rightarrow A}$ to two
parts, the channels from Alice to Bob $b_{A\rightarrow B}$ and
from Bob to Alice $b_{B\rightarrow A}$) and (ii) if loss happens,
whether it happens in the channels $b_{A\rightarrow B}$ or
$b_{B\rightarrow A}$.

\emph{Case I.} Channel loss does not happen either on
$b_{A\rightarrow B}$ or $b_{B\rightarrow A}$.

This case just like the single photon has transmitted in a
no-lossy channel. Namely, there are not any blocks except the
possible block from Bob's PBS. Case I will generate a raw key bit
with probability $P_1=\frac{RT}{2}$ as the reasons (1) Bob's PBS
blocks the special polarization with probability $\frac{1}{2}$ (2)
a raw key bit will be generated with probability $RT$ when Bob's
PBS blocks the special polarization.

As both of the loss rates in the channels $b_{A\rightarrow B}$ and
$b_{B\rightarrow A}$ are $\eta$, channel loss will not happen on
$b_{A\rightarrow B}$ and $b_{B\rightarrow A}$ with probability
$1-\eta$ respectively. So the Case I, channel loss does not happen
either on $b_{A\rightarrow B}$ or $b_{B\rightarrow A}$, will occur
with probability $P_I=(1-\eta)^2$. The raw key rate comes from
Case I is
\begin{equation}
R_{raw_1}^{AB}=P_I\cdot P_1\cdot
R_{single}=(1-\eta)^2\cdot\frac{RT}{2}\cdot R_{single}.
\end{equation}
Alice and Bob's raw key are identify in this case.

\emph{Case II.} Channel loss happens in the channel
$b_{A\rightarrow B}$, regardless of whether channel loss happens
in the channel $b_{B\rightarrow A}$ or not.

When channel loss happened in $b_{A\rightarrow B}$, no wave pulse
will pass through $b_{B\rightarrow A}$, so we combine the cases
that (II-1) channel loss happens both in the channels
$b_{A\rightarrow B}$ and $b_{B\rightarrow A}$ (II-2) channel loss
only happens in the channel $b_{A\rightarrow B}$, not in the
channel $b_{B\rightarrow A}$ to Case II. Case II will generate an
additional raw key bit with probability $P_2=RT$ as following
analysis.

Without loss of generality, we suppose the single photon Alice
sent is $|V\rangle$. After $BS_1$, the state could be described as
Eq.(1a). When it comes into Bob's site, the state evolves to
$|V\rangle_a|0\rangle_b$ with probability $T$, or
$|0\rangle_a|0\rangle_b$ with probability $R$ as the possible
pulse wave $|V\rangle_b$ lost in the channel $b_{A\rightarrow B}$.
The state $|0\rangle_a|0\rangle_b$ will not lead to any clicks, so
no raw key will be generated. But as the state
$|V\rangle_a|0\rangle_b$, the photon in path $a$ will fire
detector $D_1$ and let Alice generate a raw key bit with
probability $R$, fire detector $D_2$ with probability $T$. After
Alice announced that $D_1$ clicked, Bob would generate an
according raw key bit based on his $PBS$'s basis, i.e., state in
which basis is sent toward $D_3$. So Case II will generate an
additional raw key bit with probability $T\cdot R$ as following
analysis.

Case II will happen with probability $P_{II}=\eta$. Hence, with
the loss in channel from Alice to Bob, additional raw key bits are
generated, the totally rate of which is
\begin{equation}
R_{raw_2}^{AB}=P_{II}\cdot P_2 \cdot R_{single}=\eta\cdot RT\cdot
R_{single}.
\end{equation}
Since Bob has chosen his $PBS$'s basis randomly, his raw key bit
will be identify, and different with Alice's with equal
probability $1/2$. So both of the correct and error raw key rates
are $\frac{R_{raw_2}^{AB}}{2}$.

\emph{Case III.} Channel loss does not happen in the channel
$b_{A\rightarrow B}$, but happens in the channel $b_{B\rightarrow
A}$. Namely, a complete block is in the channel $b_{B\rightarrow
A}$ except the possible block from Bob's PBS. Case III will
generate a raw key bit with probability $P_3=RT$ as following
analysis.

We still suppose the single photon Alice sent is $|V\rangle$. If
Bob's $PBS$ basis is same with Alice's basis, Bob's $PBS$ will
send possible wave pulse $|V\rangle_b$ toward $D_3$. On one hand,
the system state evolves to $|0\rangle_a|V\rangle_b$ with
probability $R$, which means that the photon went through path
$b$, and it will be destroyed by detector $D_3$. So no pulse wave
will transmit from Bob to Alice. On the other hand, the system
state evolves to $|V\rangle_a|0\rangle_b$ with probability $T$,
which means that the photon went through path $a$, then it will
fire detectors $D_1$ and $D_2$ with probabilities $R$ and $T$
respectively. Bob will generate a raw key bit which is identify
with Alice's after she announces that $D_1$ clicked, whose
probability is $T\cdot R$.

Else if Bob's $PBS$ basis is different with Alice's basis, Bob's
$PBS$ would pass possible wave pulse $|V\rangle_b$, and send it
back to Alice. After it lost in the channel from Bob to Alice, the
system state evolves to $|0\rangle_a|0\rangle_b$ (with probability
$R$) which means that it is destroyed by the lossy channel, or
$|V\rangle_a|0\rangle_b$ (with probability $T$) which means that
it will fire $D_1$ or $D_2$ with probabilities $R$ and $T$,
respectively. Bob will generate a raw key bit which is identify
with Alice's after she announces that $D_1$ clicked, whose
probability is $T\cdot R$.

So regardless Bob's $PBS$ basis is $|V\rangle$ or $|H\rangle$,
this case will generate a raw key bit with probability $RT$. But
Alice's and Bob's bits are same and different with equal
probability $\frac{1}{2}$.

The case will happen with probability $P_{III}=(1-\eta)\cdot\eta$
as channel loss does not happen in the channel $b_{A\rightarrow
B}$ with probability $1-\eta$, happens in the channel
$b_{B\rightarrow A}$ with probability $\eta$. Hence, with the loss
in channel from Bob to Alice, additional raw key bits is
generated, the totally rate of which is
\begin{equation}
R_{raw_3}^{AB}=P_{III}\cdot P_3\cdot
R_{single}=(1-\eta)\cdot\eta\cdot RT\cdot R_{single}.
\end{equation}
Both of the same and different raw key rates are
$\frac{R_{raw_3}}{2}$.

All in all, the raw key rate is
\begin{equation}
\begin{array}{ll}
R_{raw}^{AB}=R_{raw_1}^{AB}+R_{raw_2}^{AB}+R_{raw_3}^{AB}\\
\hspace{9mm}=\frac{1+2\eta-\eta^2}{2}\cdot TR\cdot R_{single}.
\end{array}
\end{equation}
The probability of that Alice's and Bob's raw keys in a same order
are identify is
\begin{equation}
\begin{array}{ll}
P_{raw}^{AB\_same}=\frac{R_{raw_1}^{AB}+\frac{R_{raw_2}^{AB}}{2}+\frac{R_{raw_3}^{AB}}{2}}{R_{raw}}\\
\hspace{16mm}=\frac{1}{1+2\eta-\eta^2},
\end{array}
\end{equation}
the probability of that they are different is
\begin{equation}
\begin{array}{ll}
P_{raw}^{AB\_diff}=\frac{\frac{R_{raw_2}^{AB}}{2}+\frac{R_{raw_3}^{AB}}{2}}{R_{raw}}\\
\hspace{15mm}=\frac{2\eta-\eta^2}{1+2\eta-\eta^2}.
\end{array}
\end{equation}
Namely, in users' raw key pair, the error rate is
$P_{raw}^{AB\_diff}$ which should be correct by some following
classical postprocessing such as information reconciliation.

The error in users' raw key pairs will give a lot of chances to
Eve to perform some attacks. But to Eve, the first aim is that her
attacks should not be detected by the users. Following
polarization-splitting-measurement attack is one of the attacks.

\section{\label{sec:level1}Polarization-splitting-measurement attack}

Usually, we assume that Eve has unlimited technological, which is
only limited by the laws of nature. So Eve could replace the lossy
channel by a perfect quantum channel, and use the excess power for
her mischievous purposes. In this section, we first give an attack
method which can cheat the raw key bits and be concealed by the
practical lossy channel with loss rate $\frac{1}{2}$, then give
the special cheat strategies according to special loss rate range
for cheating maximal information.

In the \emph{attack method}, polarization-splitting and
measurement will be used to cheat secret information from channel
$b_{A\rightarrow B}$ (shown in fig.2). Eve first replaces the
lossy channel $b_{A\rightarrow B}$ by a perfect quantum channel.
She also has two polarizing beam splitters, $|V\rangle$ PBS
representing the bit value $0$ and $|H\rangle$ PBS representing
the bit value $1$. She randomly chooses the $|V\rangle$ or
$|H\rangle$ PBS for the $i$th order, and inserts it in front of
Bob's site.

If Eve's $i$th bit is identical with Alice's $i$th bit, the
detector $D_4$ will click with probability $R$, else if her $i$th
bit is differ to Alice's $i$th bit, the detector $D_4$ will not
click. In other words, the case that $D_4$ clicks means that Eve's
bit is identical to Alice's $i$th bit, and the case that $D_4$
does not click means that Eve is uncertain about Alice's $i$th bit
now. We recall that Alice and Bob's raw key pair will product from
these uncertain bits corresponding to the case that detector $D_4$
does not click. So Eve cannot make sure of the raw key bit.
However, Eve could easily extract the raw key bit according to
what Alice and Bob will announce in the following processing.

\begin{figure}[h] \centering
\includegraphics[scale=0.8]{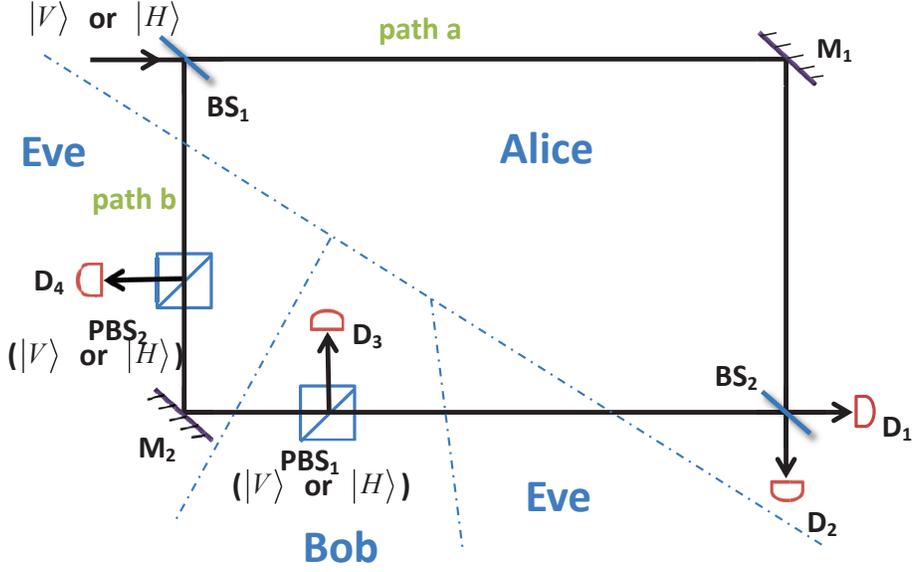}
\caption{(color online).\space The schematic of
polarization-splitting-measurement attack on counterfactual QKD.
Eve performs attack on channel $b_{A\rightarrow B}$ in front of
Bob's site. Eve replaces the lossy channel $b_{A\rightarrow B}$ by
a perfect quantum channel. Then she randomly uses $|V\rangle$
(representing $0$) or $|H\rangle$ (representing $1$) $PBS$ to
block the pulse in path $b_{A\rightarrow B}$ when her bit is
identical to Alice's, or let it pass when her bit is differ to
Alice's. What she dose just like a reasonable loss in path
$b_{A\rightarrow B}$.}
\end{figure}

Without loss of generality, we consider the case of Eve chooses
$0$, i.e., she inserts a $|V\rangle$ PBS. When Alice's bit is $0$,
two possible cases are here. (1) When Eve's detector $D_4$
clicked, the system state has been collapsed to
$|0\rangle_a|V\rangle_b$ which means Alice's bit is identical to
Eve's bit $0$, and vacuum state will go into Bob's site. (2) Else
when Eve's detector $D_4$ did not click, the system has been
collapsed to $|V\rangle_a|0\rangle_b$, and vacuum state still will
go into Bob's site. Altogether, vacuum state (i.e., nothing)
always will go into Bob's site when Eve and Alice's bits are same,
which likes the pulse in path $b$ has lost completely by the lossy
channel.

On the other hand, when Alice's bit is $1$, the pulse in path $b$
will pass Eve's $PBS_2$ completely, so the system state still is
$|\phi_1\rangle$($=
 \sqrt{R}|0\rangle_a|H\rangle_b+\sqrt{T}|H\rangle_a|0\rangle_b$) after Eve's devices.
The case is same to that Eve has done nothing, liking the ideal
setting. In the point view of Alice and Bob, all the following
processes will just like the normal processes. When $D_1$ clicks,
the corresponding bit will be chosen as a raw key bit by Alice
followed by announcing its order. Then Eve can always make sure
that Alice's raw key bit is $1$. In other words, when Eve and
Alice's bits are different, a raw key bit will be produced with
probability. And the probability will be revealed with Alice's
announcement. Since the raw key bit is generated from the inverse
of Eve's $PBS_2$'s basis, Eve can not only know the raw key bits,
but also decide its value with some probability.

Since Eve chooses bit $0$ or $1$ randomly, her bit will be same
and different with Alice's bit with probability $1/2$
respectively. The complete loss will happen when their bits are
same, and the ideal setting will happen when their bits are
different. Totally, the cheat method likes a loss of rate
$\frac{1}{2}$ happens in the channel $b_{A\rightarrow B}$. The
cheat method could be used on every photon to cheat the secret
information when $\eta=\frac{1}{2}$ and will not be detected (the
analysis will be given in the following). To other value of
$\eta$, more complex strategies should be designed for optimal
cheating.

We suppose the amount of Alice sent single photons is $n$. Using
the above attack method, Eve could simulate practical loss channel
with loss rate $0\leq\eta\leq1$ and cheat raw key bits with
following strategies.

\emph{Cheat strategy (I)} When $0\leq\eta<\frac{1}{2}$, Eve
performs the attack method on $2\eta\cdot n$ single photons
randomly, and fills the raw key orders which she has not attacked
in with random bits.

\emph{Cheat strategy (II)} When $\frac{1}{2}\leq\eta\leq1$, Eve
performs the attack method on $2(1-\eta)\cdot n$ single photons
randomly, and blocks the remaining $(2\eta-1)\cdot n$ single
photons. After Alice announced in which orders the remaining
single photons have fired detector $D_1$, she fills these raw key
orders in with random bits.

Like the loss in practical channel, what Eve did has brought some
errors to the protocol (we will analyze the details in next
section). For instance, some $D_1$'s clicks happened not only when
Alice and Bob's bits were same, but also when they were different
as long as Eve blocked the channel. However, since the error rate
is same as that brought by practical lossy channel, it will be
judged as a legal case by the protocol's detection process. The
basis reason is that, the system state under the above cheat
strategies is same to the system state transmitted from a
practical channel. We will analyze it as follows.

We suppose the photon Alice sent is $|V\rangle$. If Eve's $PBS$
past wave pulse $|V\rangle$ to Bob's site, the density matrix of
system state is
\begin{equation}
\begin{array}{ll}
\rho_1^{attack}=|\phi_0\rangle \langle\phi_0|,
\end{array}
\end{equation}
when it comes into Bob's site. If Eve's $PBS$ blocked wave pulse
$|V\rangle$, the system state is a mixed state with density matrix
\begin{equation}
\begin{array}{ll}
\rho_2^{attack}=R|0\rangle_a|0\rangle_b
\langle0|_b\langle0|_a+T|V\rangle_a|0\rangle_b\langle0|_b\langle
V|_a,
\end{array}
\end{equation}
when it comes into Bob's site.

So after the strategy (I), the system state is a mixed state with
density matrix
\begin{equation}
\begin{array}{ll}
\rho_I^{attack}
=(1-2\eta)\cdot|\phi_0\rangle \langle\phi_0|+\eta\cdot\rho_1^{attack}+\eta\cdot\rho_2^{attack}\\
\hspace{11mm}=(1-\eta)\cdot\rho_1^{attack}+\eta\cdot\rho_2^{attack},
\end{array}
\end{equation}
where $0\leq\eta<\frac{1}{2}$. After the strategy (II), the system
state is a mixed state with density matrix
\begin{equation}
\begin{array}{ll}
\rho_{II}^{attack}
=(1-\eta)\cdot\rho_1^{attack}+[(1-\eta)+(2\eta-1)]\cdot\rho_2^{attack}\\
\hspace{11mm}=(1-\eta)\cdot\rho_1^{attack}+\eta\cdot\rho_2^{attack},
\end{array}
\end{equation}
where $\frac{1}{2}\leq\eta\leq1$.

Now we analyze the system state in practical lossy channel without
the attack strategies. If the wave pulse in channel
$b_{A\rightarrow B}$ has not lost, the density matrix of the
system state is
\begin{equation}
\begin{array}{ll}
\rho_1^{loss}=|\phi_0\rangle \langle\phi_0|,
\end{array}
\end{equation}
when it come into Bob's site. If the wave pulse in channel
$b_{A\rightarrow B}$ has lost, the system state will be a mixed
state with density matrix
\begin{equation}
\begin{array}{ll}
\rho_2^{loss}=R|0\rangle_a|0\rangle_b
\langle0|_b\langle0|_a+T|V\rangle_a|0\rangle_b\langle0|_b\langle
V|_a
\end{array}
\end{equation}
when it come into Bob's site.

Since the loss rate is $\eta$ on the practical lossy channel
$b_{A\rightarrow B}$, the general system state is a mixed state
with density matrix
\begin{equation}
\begin{array}{ll}
\rho^{loss} =(1-\eta)\cdot\rho_1^{loss}+\eta\cdot\rho_2^{loss}\\
\hspace{8mm}=(1-\eta)\cdot\rho_1^{attack}+\eta\cdot\rho_2^{attack},
\end{array}
\end{equation}
when it goes into Bob's site, which is same with
$\rho_{I}^{attack}$ when $0\leq\eta<\frac{1}{2}$,
$\rho_{II}^{attack}$ when $\frac{1}{2}\leq\eta\leq1$.

So the states are same either when the protocol suffers a lossy
channel or when it is under the cheat strategies. The conclusion
is still tenable when the photon Alice sent is $|H\rangle$.
Consequently, Alice and Bob could not distinguish between the
practical lossy channel and the cheat strategies.

\section{\label{sec:level1}Secret key rate under the cheat strategies in lossy channel}

In this section, we will analyze the protocol in lossy channel
with the secret key rate $R_{QKD}$\cite{sb09,ab13}, a convenient
and commonly used quantitate measure of protocol security.

Secret key rate $R_{QKD}$ is the product of the raw key rate
$R_{raw}$ and the secret fraction $r_\infty$. The secret fraction
represents the fraction of secure bits that may be extracted from
the raw key. Formally, we have
\begin{equation}
R_{QKD}= R_{raw}\cdot r_\infty.
\end{equation}

The expression for the secret fraction
extractable\cite{sb09,reference key} using one-way classical
postprocessing reads
\begin{equation}
r_\infty=I(A;B)-\min(I_{EA},I_{EB}),
\end{equation}
where $I(A;B)$ is Alice and Bob's mutual information, $I_{EA}=
\max_{Eve}I(A;E)$, $I_{EB}= \max_{Eve}I(B;E)$. Since Alice and
Bob's each raw key pair is randomly in $\{0, 1\}$, it should be
$H(A)=H(B)=1$. We also have
$P(A=0,B=0)=P(A=1,B=1)=P_{raw}^{AB\_same}/2$,
$P(A=0,B=1)=P(A=1,B=0)=P_{raw}^{AB\_diff}/2$. Combined with
Eqs.(6) and (7), it should be that
\begin{equation}
\begin{array}{ll}
I(A;B)=H(A)+H(B)-H(A,B)\\
\hspace{12mm}=1+1+\sum_{A\in\{0,1\},B\in\{0,1\}}p(A,B)\log p(A,B)\\
\hspace{12mm}=2+2\cdot\frac{1}{2(1+2\eta-\eta^2)}\log\frac{1}{2(1+2\eta-\eta^2)}\\
\hspace{16mm}+2\cdot\frac{2\eta-\eta^2}{2(1+2\eta-\eta^2)}\log\frac{2\eta-\eta^2}{2(1+2\eta-\eta^2)}.
\end{array}
\end{equation}
Then we analyze the secret key rate under the cheat strategies (I)
and (II) respectively depend on the loss rate $\eta$ by
calculating $\min(I_{EA},I_{EB})$.

We recall that (1) Alice's raw key bits are generated from the
source single photons' bases. (2) Bob's raw key bits are generated
from his $PBS_1$'s basis, i.e., state in which basis would be sent
from $PBS_1$ toward $D_3$. (3) Eve's raw key bits are generated
from the inverse of her $PBS_2$'s basis, i.e., state in which
basis would be sent from $PBS_2$ toward Bob's site. Now we analyze
the cases in which the raw key will be cheated by Eve when she
cheats in the channel from Alice to Bob. And we still suppose the
single photon Alice sent is $|V\rangle$.

\subsection{\label{sec:level1}Secret key rate under the cheat
strategy (I) in lossy channel}

We first analyze the cases in \emph{ cheat strategy (I)}, i.e.,
the strategy with $0\leq\eta<\frac{1}{2}$. We recall \emph{Cheat
strategy (I)}: When $0\leq\eta<\frac{1}{2}$, Eve performs the
attack method on $2\eta\cdot n$ single photons randomly, and fills
the raw key orders which she has not attacked in with random bits.
We divide the cases with elements (i) whether Eve performs the
attack method or not and (ii) if Eve performs the attack method,
whether her $PBS$ basis is same with Alice's basis or not.

\emph{Cheated raw key I.} The cheated raw key when Eve does not
perform the attack method.

For $0\leq\eta<\frac{1}{2}$, Eve does not perform the attack
method on $(1-2\eta)\cdot n$ source single photons, in which raw
key bits will be generated as the \emph{case I} and \emph{case
III} (shown in Sec.III). Due to that the loss rate in the channel
$b_{A\rightarrow B}$ is $\eta$, \emph{case I} will happen with
probability $(1-2\eta)\cdot(1-\eta)$, \emph{case III} will happen
with probability $(1-2\eta)\cdot\eta$. The totally rate of these
raw key is
\begin{equation}
\begin{array}{ll}
R_{raw_1}^{E}=
((1-2\eta)\cdot(1-\eta)\cdot P_1+(1-2\eta)\cdot\eta\cdot P_3)\cdot R_{single}\\
\hspace{11mm}=\frac{1-\eta-2\eta^2}{2}\cdot RT\cdot R_{single}.
\end{array}
\end{equation}
Eve will guess these raw key bits, so the correct probability is
$\frac{1}{2}$. So compared to Alice's and Bob's raw keys, both of
Eve's same and different raw key rates in this case are
\begin{equation}
\begin{array}{ll}
R_{raw_1}^{EA\_same}=R_{raw_1}^{EA\_diff}=R_{raw_1}^{EB\_same}=R_{raw_1}^{EB\_diff}\\
\hspace{17mm}=\frac{1-\eta-2\eta^2}{4}\cdot RT\cdot R_{single}.
\end{array}
\end{equation}

\emph{Cheated raw key II.}  The cheated raw key when Eve performs
the attack method, and her $PBS$ basis is same with Alice's basis.

When Eve's $PBS$ basis is same with Alice's basis (namely, Eve's
$PBS$ will send wave pulse $|V\rangle$ toward $D_4$), raw key bits
will be generated as the \emph{case II}(shown in Sec.III). It will
happen with probability $\eta$. So the totally rate of these raw
key is
\begin{equation}
\begin{array}{ll}
R_{raw_2}^{E}=\eta\cdot P_2\cdot R_{single}\\
\hspace{11mm}=\eta\cdot RT\cdot R_{single}.
\end{array}
\end{equation}
Since Eve always generates her raw key bit as the inverse of her
$PBS_2$'s basis, all her raw key bits are different to Alice's,
and different to Bob's with probability $\frac{1}{2}$. Compared to
Alice's raw key, Eve's same and different raw key rates are
\begin{equation}
\begin{array}{ll}
R_{raw_2}^{EA\_same}=0,\\
R_{raw_2}^{EA\_diff}=\eta\cdot RT\cdot R_{single},
\end{array}
\end{equation}
respectively. Compared to Bob's raw key, Eve's same and different
raw key rates are
\begin{equation}
\begin{array}{ll}
R_{raw_2}^{EB\_same}=R_{raw_2}^{EB\_diff}=\frac{\eta}{2}\cdot
RT\cdot R_{single},
\end{array}
\end{equation}

\emph{Cheated raw key III.} The cheated raw key when Eve performs
the attack method, and her $PBS$ basis is different with Alice's
basis.

When Eve's $PBS$ basis is different with Alice's basis, Eve's
$PBS$ will send wave pulse $|V\rangle$ toward Bob's site. It will
happen with probability $\frac{1}{2}\cdot2\eta=\eta$. And raw key
bits will be generated as the \emph{case I} and \emph{case III}.
Due to that the loss rate in channel from Bob to Alice is $\eta$,
\emph{case I} will happen with probability $\eta\cdot(1-\eta)$,
\emph{case III} will happen with probability $\eta\cdot\eta$. So
the totally rate of these raw key is
\begin{equation}
\begin{array}{ll}
R_{raw_3}^{E}=[\eta\cdot(1-\eta)\cdot P_1+\eta\cdot\eta\cdot P_3]\cdot R_{single}\\
\hspace{11mm}=\frac{\eta+\eta^2}{2}\cdot RT\cdot R_{single}.
\end{array}
\end{equation}
Since Eve always generates her raw key bit as the inverse of her
$PBS_2$'s basis, all her raw key bits are identify with Alice's.
Compared to Alice's raw key, Eve's same and different raw key
rates are
\begin{equation}
\begin{array}{ll}
R_{raw_3}^{EA\_same}=\frac{\eta+\eta^2}{2}\cdot RT\cdot R_{single},\\
R_{raw_3}^{EA\_diff}=0.
\end{array}
\end{equation}
Compared to Bob's raw key, Eve's same and different raw key rates
are
\begin{equation}
\begin{array}{ll}
R_{raw_3}^{EB\_same}=\frac{\eta\cdot(1-\eta)}{2}\cdot RT\cdot R_{single},\\
R_{raw_3}^{EB\_diff}=\eta\cdot\eta\cdot RT\cdot R_{single}.
\end{array}
\end{equation}

All in all, the raw key rate Eve cheated is
\begin{equation}
\begin{array}{ll}
R_{raw}^{E}=R_{raw_1}^{E}+R_{raw_2}^{E}+R_{raw_3}^{E}\\
\hspace{9mm}=\frac{1+2\eta-\eta^2}{2}RT\cdot R_{single},
\end{array}
\end{equation}
which is same as users' raw key rate. The probabilities of that
Eve and Alice's raw key bits are same and different are
\begin{equation}
\begin{array}{ll}
P_{raw}^{EA\_same}=\frac{P_{raw_1}^{EA\_same}+P_{raw_2}^{EA\_same}+P_{raw_3}^{EA\_same}}{R_{raw}^{E}}\\
\hspace{16mm}=\frac{1+\eta}{2(1+2\eta-\eta^2)},\\
\end{array}
\end{equation}
\begin{equation}
\begin{array}{ll}
P_{raw}^{EA\_diff}=\frac{P_{raw_1}^{EA\_diff}+P_{raw_2}^{EA\_diff}+P_{raw_3}^{EA\_diff}}{R_{raw}^{E}}\\
\hspace{15mm}=\frac{1+3\eta-2\eta^2}{2(1+2\eta-\eta^2)}.
\end{array}
\end{equation}
The probabilities of that Eve and Bob's raw key bits are same and
different are
\begin{equation}
\begin{array}{ll}
P_{raw}^{EB\_same}=\frac{P_{raw_1}^{EB\_same}+P_{raw_2}^{EB\_same}+P_{raw_3}^{EB\_same}}{R_{raw}^{E}}\\
\hspace{16mm}=\frac{1+3\eta-4\eta^2}{2(1+2\eta-\eta^2)},\\
\end{array}
\end{equation}
and
\begin{equation}
\begin{array}{ll}
P_{raw}^{EB\_diff}=\frac{P_{raw_1}^{EB\_diff}+P_{raw_2}^{EB\_diff}+P_{raw_3}^{EB\_diff}}{R_{raw}^{E}}\\
\hspace{15mm}=\frac{1+\eta+2\eta^2}{2(1+2\eta-\eta^2)}.
\end{array}
\end{equation}.
In fact, Eve's error rate will not be larger than $50\%$ by using
a simple way\cite{reverse bit}.

Similar to the calculation of $I(A;B)$, combined with Eqs.(27-30)
it should be
\begin{equation}
\begin{array}{ll}
I(E;A)^{i}=H(E)+H(A)-H(E,A)\\
\hspace{14mm}=1+1+\sum_{E\in\{0,1\},A\in\{0,1\}}p(E,A)\log p(E,A)\\
\hspace{14mm}=2+2\cdot\frac{1+\eta}{4(1+2\eta-\eta^2)}\log\frac{1+\eta}{4(1+2\eta-\eta^2)}\\
\hspace{17mm}+2\cdot\frac{1+3\eta-2\eta^2}{4(1+2\eta-\eta^2)}\log\frac{1+3\eta-2\eta^2}{4(1+2\eta-\eta^2)},
\end{array}
\end{equation}
and
\begin{equation}
\begin{array}{ll}
I(E;B)^{i}=H(E)+H(B)-H(E,B)\\
\hspace{14mm}=1+1+\sum_{E\in\{0,1\},B\in\{0,1\}}p(E,B)\log p(E,B)\\
\hspace{14mm}=2+2\cdot\frac{1+3\eta-4\eta^2}{4(1+2\eta-\eta^2)}\log\frac{1+3\eta-4\eta^2}{4(1+2\eta-\eta^2)}\\
\hspace{17mm}+2\cdot\frac{1+\eta+2\eta^2}{4(1+2\eta-\eta^2)}\log\frac{1+\eta+2\eta^2}{4(1+2\eta-\eta^2)},
\end{array}
\end{equation}

Then the secret fraction is
\begin{equation}
\begin{array}{ll}
r_\infty^{i}=I(A;B)-\min(I_{EA}^i,I_{EB}^i),
\end{array}
\end{equation}
where $0\leq\eta<\frac{1}{2}$.

For simpleness, we set $R=T=\frac{1}{2}$. Then secret key rate is
\begin{equation}
\begin{array}{ll}
R_{QKD}= R_{raw}\cdot r_\infty^{i}\\
\hspace{11mm}=\frac{1+2\eta-\eta^2}{8}\cdot r_\infty^{i}\cdot
R_{single},
\end{array}
\end{equation}
where $0\leq\eta<\frac{1}{2}$.

\subsection{\label{sec:level1}Secret key rate under the cheat
strategy (II) in lossy channel}

Now we analyze the cases in which the raw key will be cheated by
Eve using \emph{cheat strategy (II)}, i.e., the strategy with
$\frac{1}{2}\leq\eta\leq1$. We recall \emph{Cheat strategy (II)}:
When $\frac{1}{2}\leq\eta\leq1$, Eve performs the attack method on
$2(1-\eta)\cdot n$ single photons randomly, and blocks the
remaining $(2\eta-1)\cdot n$ single photons. After Alice announced
in which orders the remaining single photons have fired detector
$D_1$, she fills these raw key orders in with random bits.

In the strategy, the attack is performed with probability
$2(1-\eta)$ replacing the probability $2\eta$ in \emph{cheat
strategy (I)}. So the amount of raw key rate generated by the
attack is $\frac{1-\eta}{\eta}\cdot
(R_{raw_2}^{E}+R_{raw_3}^{E})$.

In addition, Eve blocks the remaining $(2\eta-1)\cdot n$ wave
pulses in the channel $b_{A\rightarrow B}$ followed by guessing
the possible raw key bits. This just likes \emph{case II}. It will
generate raw key bits whose amount is $(2\eta-1)\cdot P_2\cdot
R_{single}$. And both of the probabilities of them are same and
different with Alice (and Bob's) are $\frac{1}{2}$.

Hence, the raw key rate is
\begin{equation}
\begin{array}{ll}
R_{raw}^{E} =[\frac{1-\eta}{\eta}
\cdot(R_{raw_2}^{E}+R_{raw_3}^{E})+(2\eta-1)\cdot P_2]\cdot R_{single}\\
\hspace{9mm}=\frac{1+2\eta-\eta^2}{2}RT\cdot R_{single},
\end{array}
\end{equation}
which is same as users' raw key rate. The probabilities of that
Eve's and Alice's raw key bits are same and different are
\begin{equation}
\begin{array}{ll}
P_{raw}^{EA\_same} =\frac{\frac{1-\eta}{\eta}\cdot
(R_{raw_2}^{EA\_same}+R_{raw_3}^{EA\_same})+\frac{(2\eta-1)}{2}\cdot P_2}{R_{raw}^{E}}\\
\hspace{16mm}=\frac{2\eta-\eta^2}{1+2\eta-\eta^2},\\
\end{array}
\end{equation}
and
\begin{equation}
\begin{array}{ll}
P_{raw}^{EA\_diff} =\frac{\frac{1-\eta}{\eta}\cdot
(R_{raw_2}^{EA\_diff}+R_{raw_3}^{EA\_diff})+\frac{(2\eta-1)}{2}\cdot
P_2}{R_{raw}^{E}}\\
\hspace{16mm}=\frac{1}{1+2\eta-\eta^2}.
\end{array}
\end{equation}
The probabilities of that Eve's and Bob's raw key bits are same
and different are
\begin{equation}
\begin{array}{ll}
P_{raw}^{EB\_same} =\frac{\frac{1-\eta}{\eta}\cdot
(R_{raw_2}^{EB\_same}+R_{raw_3}^{EB\_same})+\frac{(2\eta-1)}{2}\cdot P_2}{R_{raw}^{E}}\\
\hspace{16mm}=\frac{1-\eta+\eta^2}{1+2\eta-\eta^2},
\end{array}
\end{equation}
and
\begin{equation}
\begin{array}{ll}
P_{raw}^{EB\_diff} =\frac{\frac{1-\eta}{\eta}\cdot
(R_{raw_2}^{EB\_diff}+R_{raw_3}^{EB\_diff})+\frac{(2\eta-1)}{2}\cdot
P_2}{R_{raw}^{E}}\\
\hspace{16mm}=\frac{3\eta-2\eta^2}{1+2\eta-\eta^2}.
\end{array}
\end{equation}

Then we have
\begin{equation}
\begin{array}{ll}
I(E;A)^{ii}=H(E)+H(A)-H(E,A)\\
\hspace{15mm}=1+1+\sum_{E\in\{0,1\},A\in\{0,1\}}p(E,A)\log p(E,A)\\
\hspace{15mm}=2+2\cdot\frac{2\eta-\eta^2}{2(1+2\eta-\eta^2)}\log\frac{2\eta-\eta^2}{2(1+2\eta-\eta^2)}\\
\hspace{18mm}+2\cdot\frac{1}{2(1+2\eta-\eta^2)}\log\frac{1}{2(1+2\eta-\eta^2)},
\end{array}
\end{equation}
and
\begin{equation}
\begin{array}{ll}
I(E;B)^{ii}=H(E)+H(B)-H(E,B)\\
\hspace{15mm}=1+1+\sum_{E\in\{0,1\},B\in\{0,1\}}p(E,B)\log p(E,B)\\
\hspace{15mm}=2+2\cdot\frac{1-\eta+\eta^2}{2(1+2\eta-\eta^2)}\log\frac{1-\eta+\eta^2}{2(1+2\eta-\eta^2)}\\
\hspace{18mm}+2\cdot\frac{3\eta-2\eta^2}{2(1+2\eta-\eta^2)}\log\frac{3\eta-2\eta^2}{2(1+2\eta-\eta^2)},
\end{array}
\end{equation}

Then the secret fraction is
\begin{equation}
\begin{array}{ll}
r_\infty^{ii}=I(A;B)-\min(I_{EA}^{ii},I_{EB}^{ii}),
\end{array}
\end{equation}
where $\frac{1}{2}\leq\eta\leq1$.

For simpleness, we set $R=T=\frac{1}{2}$. Then secret key rate is
\begin{equation}
\begin{array}{ll}
R_{QKD}= R_{raw}\cdot r_\infty^{ii}\\
\hspace{11mm}=\frac{1+2\eta-\eta^2}{8}\cdot r_\infty^{ii}\cdot
R_{single},
\end{array}
\end{equation}
where $\frac{1}{2}\leq\eta\leq1$.

\subsection{\label{sec:level1}Discuss of the secret key rate}

\begin{figure}[h]
\centering
\includegraphics[scale=0.7]{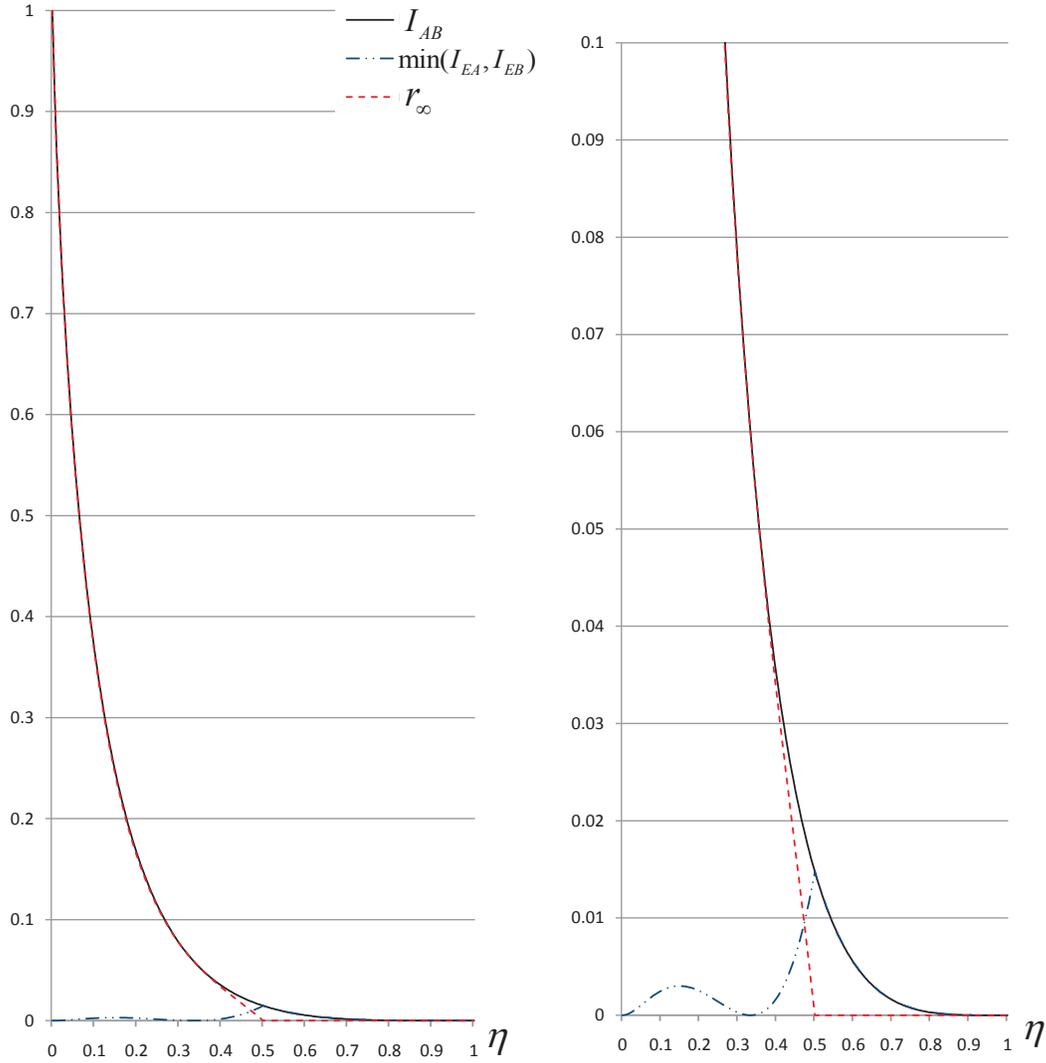}
\caption{(color online).\space $I(A;B)$ is Alice's and Bob's
mutual information. $min(I(E;A), I(E;B))$ is the minimum of Eve's
and Alice's, Eve's and Bob's mutual information. $r_\infty$ is the
secret fraction. They are given compared to loss rate $\eta$. The
left figure is the whole show of them. In the right figure, the
ordinate scale is magnified.}
\end{figure}
\begin{figure}[h] \centering
\includegraphics[scale=0.7]{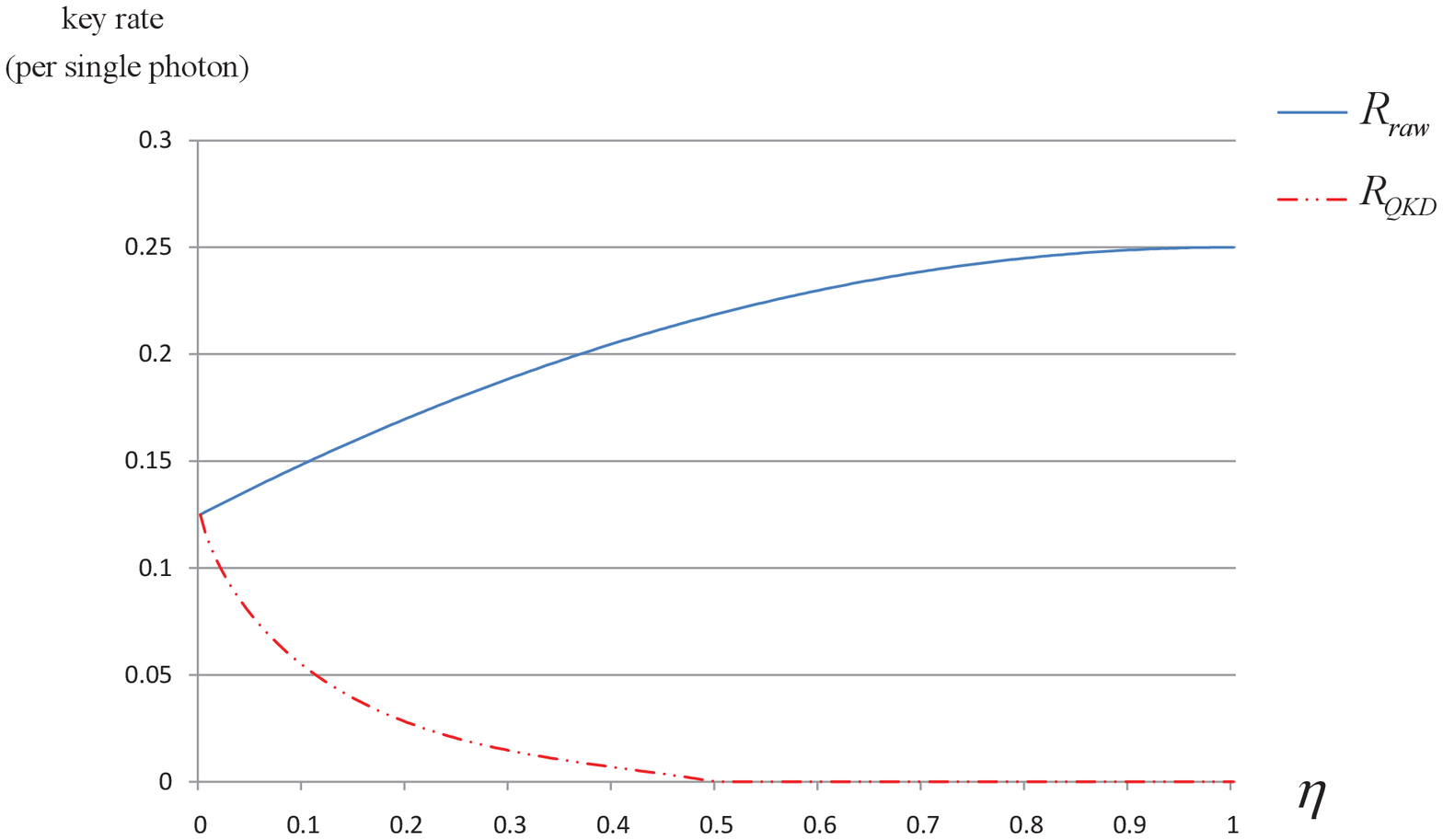}
\caption{(color online).\space The raw key rate $R_{raw}$ and the
secret key rate $R_{QKD}$ compared to loss rate $\eta$. Here the
key rate is the key bit rate generated by one single photon, and
we set $T=R=1/2$. The left figure is the whole show of them. In
the right figure, the ordinate scale is magnified.}
\end{figure}
Fig.3 shows Alice and Bob's mutual information $I(A;B)$, the
minimum of Eve's and Alice's, Eve's and Bob's mutual information
$min(I(E;A),I(E;B))$ when Eve uses the cheat strategies \emph{(I)}
and \emph{(II)}, and the secret fraction
$r_\infty$($=I(A;B)-min(I(E;A),I(E;B))$) compared to the loss rate
$\eta$. It indicates that $r_\infty=0$ when
$\frac{1}{2}\leq\eta\leq1$ under the cheat strategies.

We explain something about the data. When
$\frac{1}{2}\leq\eta\leq1$, $min(I(E;A),I(E;B))= I(E;A)$, which is
monotonic. But when $0\leq\eta<\frac{1}{2}$, it will be
$min(I(E;A),I(E;B))= I(E;B)$, which is not monotonic. Specially,
when $\eta=\frac{1}{3}$, minimal value $I(E;B)=0$ is here with
$P_{raw}^{EB\_same}=P_{raw}^{EB\_diff}$. The reason is that
information entropy is non-negative. With the increasing of
disparity between $\eta$ and the special value $\frac{1}{3}$, the
disparity between $P_{raw}^{EB\_same}$ and $P_{raw}^{EB\_diff}$
increases, consequently, $I(E;B)$ increases. (Also see
\cite{reverse bit})

Fig.4 shows the counterfactual QKD's raw key rate $R_{raw}$ and
the secret key rate $R_{QKD}$ compared to the loss rate $\eta$. It
indicates that $R_{raw}$ increases with the increasing of $\eta$,
$R_{QKD}$ decreases with the increasing of $\eta$. Specially,
$R_{QKD}$ will be equal to $0$ when $\frac{1}{2}\leq\eta\leq1$
under the cheat strategies, which means the protocol is insecure.

As QKD applications, they usually need to distribute secret
information over long distance, so the high loss rate of channel
is inevitable. For instance, let us assume that the transmission
line is a fiber-based channel, which is always slightly lossy
(about $0.2$ dB/km). If we want to use the cryptographic system
over reasonable distances, say up to $15$ km, transmission losses
will be as high as $3$dB, or about $50\%$. Then Eve could cheat
all the secret information using the cheat strategies proposed
without leaving any clues.

\section{\label{sec:level1}Conclusion}

In conclusion, we pointed out that counterfactual
cryptography\cite{n09} is insecure in practical high lossy
channel. We proposed a polarization-splitting-measurement attack
and analyzed the secret key rate in lossy channel. The analysis
indicates that the protocol is insecure when the loss rate of the
channel from Alice to Bob is up to $50\%$. Since the attack's
effect just likes a loss channel, it is invisible to the
protocol's participants. Maybe the security flaw could be overcome
by using nonorthogonal states as BB84 QKD\cite{bb84}, but the
protocol will be more complex and lower efficient.

\ack We are very grateful to Professor Horace P. Yuen for
encouragement. This work is supported by NSFC (Grant Nos.
61300181, 61272057, 61202434, 61170270, 61100203, 61121061,
61370188, and 61103210), Beijing Natural Science Foundation (Grant
No. 4122054), Beijing Higher Education Young Elite Teacher
Project, China scholarship council.

\newpage 
\bibliography{apssamp}

\begin{thebibliography}{00}
\bibitem{reference key} It will be
$I_{EB}<I_{EA}$ (or $I_{EB}>I_{EA}$) when $\eta$'s value is in
some ranges. It means that in the counterfactual QKD, Bob's (or
Alice's) raw key should be chosen as a reference raw key followed
by classical postprocessing for more secure against Eve's attack.
So the mutual information Eve obtained is $I_{EB}$ (or $I_{EA}$),
i.e., $\min(I_{EA},I_{EB})$.
\bibitem{reverse bit} It will be
$R_{raw}^{EA\_same}< R_{raw}^{EA\_diff}$ ($R_{raw}^{EB\_same}<
R_{raw}^{EB\_diff}$) when $\eta$'s value is in some ranges. In
practical attack, Eve should reverse all her cheated bits for
decreasing the error rate to less than $50\%$ based on the value
of $\eta$. Then the information she obtained is corresponding to
the result calculated from information theory. And what she did
will not affect the calculation in information theory, so we have
not emphasized this in the rest of this paper.
\end{thebibliography}

\end{document}